\documentclass[proceedings]{aofa}

\usepackage[utf8]{inputenc}
\usepackage{subfigure}

\usepackage{amsmath}
\usepackage{amsfonts}
\usepackage{tikz}
\newtheorem{lemma}{Lemma}
\newtheorem{theorem}{Theorem}
\newtheorem{corollary}{Corollary}
\newcommand{\tail}[1]{\overline{#1}}
\newcommand{\ttail}[1]{\tail{\tail{#1}}}
\newcommand{\tttail}[1]{\tail{\ttail{#1}}}

\author[P.V Poblete and A. Viola]{Patricio V. Poblete\addressmark{1}\thanks{
Supported in part by NIC Chile
}\and Alfredo Viola\addressmark{2}
\thanks{
This work has been partially supported by Project
CSIC I+D "Combinatoria Anal\'{\i}tica y aplicaciones en
criptograf\'{\i}a, comunicaciones y recuperaci\'on de la
informaci\'on", fondos 2015-2016.}
}

\title{Robin Hood Hashing \emph{really} has constant average search cost and variance in full tables}

\address{\addressmark{1}Dept. of Computer Science, University of Chile, Chile\\
\addressmark{2}Universidad de la Rep\'ublica, Uruguay}


\begin{document}

\maketitle
\begin{abstract}
Thirty years ago, the Robin Hood collision resolution strategy was introduced for open addressing hash tables, and a recurrence equation was found for the distribution of its search cost. Although this recurrence could not be solved analytically, it allowed for numerical computations that, remarkably, suggested that the variance of the search cost approached a value of $1.883$ when the table was full. Furthermore, by using a non-standard mean-centered search algorithm, this would imply that searches could be performed in expected constant time even in a full table.

In spite of the time elapsed since these observations were made, no progress has been made in proving them.
In this paper we introduce a technique to work around the intractability of the recurrence equation by solving instead an associated differential equation. While this does not provide an exact solution, it is sufficiently powerful to prove a bound for the variance, and thus obtain a proof that the variance of Robin Hood is bounded by a small constant for load factors arbitrarily close to 1. As a corollary, this proves that the mean-centered search algorithm runs in expected constant time.

We also use this technique to study the performance of Robin Hood hash tables under a long sequence of insertions and deletions, where deletions are implemented by marking elements as {\em deleted}. We prove that, in this case, the variance is bounded by $1/(1-\alpha)+O(1)$, where $\alpha$ is the load factor.

To model the behavior of these hash tables, we use a unified approach that can be applied also to study the First-Come-First-Served and Last-Come-First-Served collision resolution disciplines, both with and without deletions.

\end{abstract}

\section{Introduction}
In 1986,  Celis {\it{et al}} \cite{CelisT,Celis} introduced the Robin Hood collision resolution strategy for open addressing hash tables. Under this discipline, collisions are decided in favor of the element that is farthest from its home location.
While this does not change the expected search cost, it turns out to have a dramatic effect on its {\em variance}. In effect, unlike other disciplines where the variance tends to infinity as the table becomes full, the variance of Robin Hood seems to remain constant, and very small. This fact, conjectured from numerical computations, has not been proved in the years since it was observed, and is the main focus of our work. This problem has been hard to solve because the distribution of the search cost obeys a nonlinear recurrence equation for which no successful line of attack has been found.

To show the kind of recurrence involved, we quote now Theorem 3.1 from \cite{CelisT} (our notation will be slightly different):

\noindent{\bf Theorem 3.1} {\em In the asymptotic model for an infinite Robin Hood hash table with load factor $\alpha$ ($\alpha<1$), the probability $p_i(\alpha)$ that a record is placed in the $i$-th or further position in its probe sequence is equal to}
\begin{equation}
p_1(\alpha) = 1, \quad 
p_{i+1}(\alpha) = 1-\left( \frac{1-\alpha}{\alpha} \right) \left( e^{\alpha(p_1(\alpha)+\cdots+p_i(\alpha))}  \right).
\end{equation}

They then go on to define another function $r_i(\alpha)=\alpha(p_i(\alpha)+\cdots+p_{\infty}(\alpha))$, in terms of which the variance can be expressed as
\begin{equation}
V(\alpha) = \frac{2}{\alpha}\sum_{i=1}^{\infty} r_i(\alpha) + \frac{\ln(1-\alpha)}{\alpha}-\frac{\ln^2(1-\alpha)}{\alpha^2}.
\end{equation}

They show that $r_i(\alpha)$satisfies the following recurrence equation:
\begin{equation}
\label{eq:Celisr}
r_i(\alpha)-r_{i+1}(\alpha) = 1-e^{-r_i(\alpha)}
\end{equation}
with $r_1(\alpha)=-\ln(1-\alpha)$.
By leaving the ``$(\alpha)$'' implicit and using the $\Delta$ operator (defined as $\Delta r_i=r_{i+1}-r_i$), this can be rewritten as
$\Delta  r_i = f(r_i)$
where $f$ is the function $f(x)=-1+e^{-x}$.

This seemingly simpler equation has, nonetheless, so far remained unsolved.

In this paper, we will introduce a technique applicable to equations of this form, and we will use it first to prove a bound on the variance of Robin Hood hashing. Then we will use it to study another recurrence equation of the same type arising from the problem of hashing with deletions.

\section{Modeling hashing algorithms}

In this paper we will study the search cost of a
random element in a hash table, using the \emph{random probing model}.
This is an open addressing
hashing scheme in which collisions are resolved by additional probes
into the table.  The sequence of these probes are considered to be
random and depends only on the value of the key. The difference with
uniform probing is that positions may be repeated in this sequence.
We use the {\em asymptotic model} for a hash table with load factor $\alpha$ \cite{guibas1976analysis,Guibas:1978:AHT:322092.322096,Celis,Mit}, where we assume that
the number of keys $n$ and the table size $m$
both tend to infinity, maintaining constant their ratio $\alpha = n/m$.

Each element has associated with it
an infinite probe sequence consisting of
i.i.d.\ integers uniformly distributed over
$\{ 0,\ldots, m-1\}$, representing the
consecutive places of probes for that element.
The probe sequence for element $x$ is
denoted by $h_1(x), h_2(x), h_3(x), \ldots$.
Elements are inserted sequentially into the table.
If element $x$ is placed in position $h_j(x)$,
then we say that element $x$ has age $j$,
as it requires $j$ probes to reach the element in
case of a search.
When an element $x$ of age $j$ and
an element $y$ of age $k$ compete for the same slot ($h_j(x)=h_k(y)$),
a collision resolution strategy is needed.

In the standard method, a collision is resolved in favor of the incumbent key, so the incoming key continues probing to its next location.
We call this a First-Come-First-Served (FCFS) collision resolution discipline.
Several authors \cite{Brent,Amble,GM} observed that a collision could be
resolved in favor of {\it{any}} of the keys involved, and used this
additional degree of freedom to decrease the expected search time in the
table.

Celis {\it{et al}} \cite{CelisT,Celis} were the first to
observe that
collisions could be resolved having instead {\it{variance reduction}} as a goal.
They defined the Robin Hood (RH) heuristic, in which each collision occurring
during an insertion is resolved in
favor of the key that is farthest away from its home location (i.e., oldest in terms of {\em age}).
Later, Poblete and Munro \cite{LCFS} defined the Last-Come-First-Served
heuristic,
where collisions are resolved in favor of the {\em incoming} key. 

In both cases,
the variance is reduced, and this can be used to speed up searches by
replacing
the standard search algorithm by a {\em mean-centered} one that first
searches in the vicinity of where we would expect the element to have
{\em drifted} to, rather than in its initial probe location.
This {\em mean-centered} approach was introduced in
\cite{CelisT} (and called ``organ-pipe search'') to speed up successful searches 
in the Robin Hood heuristic, 
with expected cost bounded by the standard deviation of this random variable. 
Numerical computations in \cite{CelisT} suggest that for
full tables the variance of the search cost for RH is constant, but no formal proof is given. 

In this paper we formally settle this conjecture, by proving that this is
in fact the case, and give an explicit upper bound (although not as tight
as the numerical results seem to suggest). As a consequence we prove
that the mean-centered searching algorithm in \cite{CelisT} has constant
expected cost for full tables.

In section \ref{conborrados} we extend this approach
to perform the analysis of hashing with deletions. 
Deletions in open addressing hash tables are often handled by marking the
cells as {\em deleted} instead of {\em empty}, because otherwise the
search algorithm might fail to find some of the keys.
The space used by deleted cells may be reused by subsequent
insertions.
Intuitively, search times should deteriorate as tables become
contaminated with deleted cells and, as Knuth\cite{Knuth3} points out,
in the long run the average successful search time should approach the
average {\em unsucessful} search time.

In this paper we analize the effect of a long sequence of insertions
and deletions in the asymptotic regime ($\alpha$-full tables with $0\leq
\alpha < 1$)
and prove a bound for the variance of RH with deletions that is close to numerical results.

There is an alternative algorithm designed to keep variance low in the presence of deletions. This method marks cells as deleted, but keeps the key values (these cells are called {\em tombstones}).
In this paper we do not study the algorithm with tombstones.
We note that \cite{Mit} derives equations for this algorithm, but only obtains
numerical solutions.

\section{Analysis without deletions}
\label{sinborrar}
To analyze the cost of searching for a random element, we begin by presenting a general framework, 
based on the one used in \cite{cunto1988two}. 
This framework applies also to FCFS and LCFS, but in this paper
we use it to analyze RH, which has been a long standing open problem.
As stated before, we use the asymptotic model for a hash table with load factor $\alpha$ and random probing.

Under this model, if collisions are resolved without ``looking ahead" in the table, the cost of inserting a random element is 1 plus a random variable that follows
a geometric distribution with parameter $1-\alpha$, and therefore its expected cost is $1/(1-\alpha)$, independently of the collision resolution discipline used.

Let $p_i(\alpha)$ be the probability that a randomly chosen key has
age $i$ when the table has load factor $\alpha$.

Suppose we insert a new element.
Depending on the insertion discipline used, a number of keys will change
locations and therefore increase their ages as a consequence of the arrival of the new element.
Let us call $t_i(\alpha)$ the expected number of probes made by keys of age $i$ during the course of the insertion.
It is easy to see that
\begin{equation}
t_1(\alpha) = 1, \quad
\sum_{i\ge 1} t_i(\alpha) =\frac {1}{1-\alpha}.
\end{equation}
Before the insertion, the expected number of keys of age $i$ is
$\alpha m p_i(\alpha)$.
After the insertion, it is
\begin{equation}\label{eq:ins}
(\alpha m+1)p_i(\alpha+\frac{1}{m}) =
\alpha m p_i(\alpha) + t_i(\alpha) - t_{i+1}(\alpha)
\end{equation}
If we write $\Delta\alpha = 1/m$ and $q_i(\alpha)=\alpha p_i(\alpha)$, this equation becomes
\begin{equation}
\frac{q_i(\alpha+\Delta\alpha)-q_i(\alpha)}{\Delta\alpha}
= t_i(\alpha) - t_{i+1}(\alpha)
\end{equation}
and, as $\Delta\alpha \rightarrow 0$ (i.e. $m \rightarrow \infty$),
\begin{equation}
\label{eq:diffi}
\partial_{\alpha}q_i(\alpha) = t_i(\alpha) - t_{i+1}(\alpha),
\end{equation}
where $\partial_{\alpha}$ denotes a derivative with respect to $\alpha$, and with the initial condition $q_i(0)=0$.

We introduce a notation that we will use throughout the paper. For any sequence $a_i$ we define its {\em tail} $\tail{a}_i$ as
\begin{equation}
\tail{a}_i = \sum_{j \ge i} a_j.
\end{equation}

Using this, equation (\ref{eq:diffi}) can be rewitten as

\begin{equation}
\label{eq:diffi1}
\partial_{\alpha} \tail{q}_i(\alpha) = t_i(\alpha).
\end{equation}

We note that this equation is valid for all three
collision resolution strategies, and it generalizes
formula (10) in \cite{Mit}, where it is proved only
for RH.

The mean of the search cost can be obtained using the tail notation, as
\begin{equation}\label{eq:tailE}
\mu_{\alpha}=\ttail{p}_1(\alpha)=\frac{1}{\alpha}\ttail{q}_1(\alpha)
\end{equation}
and the variance as
\begin{equation}\label{eq:tailV}
\sigma_{\alpha}^2 = 2\tttail{p}_1(\alpha) - \mu_{\alpha} - \mu_{\alpha}^2
= \frac{2}{\alpha} \tttail{q}_1(\alpha) - \mu_{\alpha} - \mu_{\alpha}^2
\end{equation}

We note that we can already compute the expected search cost, without needing to know the exact form of the function $t_i(\alpha)$.
Taking tails in both sides of (\ref{eq:diffi1}), we have
$\partial_{\alpha} \ttail{q}_i(\alpha) =
\tail{t}_i(\alpha)$.

Now setting $i=1$ and using (\ref{eq:tailE}), we obtain
$\partial_{\alpha} (\alpha\mu_{\alpha}) = \frac{1}{1-\alpha}$,
and from this we obtain
\begin{equation}
\label{eq:mu}
\mu_{\alpha} = \frac{1}{\alpha}\ln{\frac{1}{1-\alpha}}
\end{equation}
independently of the collision resolution discipline used.

The fact that the mean search cost is independent of the collision resolution discipline used does not necessarily carry over to higher moments or to the distribution of the search cost. To compute them, we need to know the $t_i(\alpha)$ for the specific discipline.

For RH, a key will be forced to try its $(i+1)$st probe location or higher each time there is a collision between an incoming key of age $i$ or higher and another key in the table that is also of age $i$ or higher. Therefore, and leaving the ``$(\alpha)$'' implicit, to simplify notation, we have:

\begin{equation}
\label{eq:qiti}
\tail{t}_{i+1} = \tail{t}_i  \tail{q}_i
\end{equation}

Together with equation (\ref{eq:diffi}) this implies
$\partial_{\alpha} \tail{q}_i = (1-\tail{q}_i)
\partial_{\alpha} \ttail{q}_i$.
Then, after integrating both sides of the equation
we have
$\ln \frac{1}{1-\tail{q}_i} = \ttail{q}_i$
from where we obtain
$\tail{q}_i = 1 - e^{-\ttail{q}_i}$.
Moreover, by expressing $\tail{q}$ as the difference
of two $\ttail{q}$, we arrive at
\begin{theorem}\label{theorem:RH}

Under the asymptotic model for an infinite hash table with random probing, and Robin Hood collision resolution discipline, the double tail of the probability distribution of the search cost of a random element satisfies the recurrence
\begin{equation}\label{eq:recurRH}
\ttail{q}_{i+1} = \ttail{q}_i - 1 + e^{-\ttail{q}_i}
\end{equation}
with the initial condition $\ttail{q}_1=\ln \frac{1}{1-\alpha}$. \qed
\end{theorem}

This is exactly equation (\ref{eq:Celisr}) that we quoted from \cite{CelisT}, but we obtained it through a completely different derivation.
As we mentioned before, numerical computations performed in \cite{Celis} indicate that as
$\alpha \rightarrow 1$, the variance converges to a small constant, approximately equal to $1.883$.

\subsection{Bounding the variance of RH}\label{BoundingRH}

Since we are interested in the behavior of the method as $\alpha \rightarrow 1$, we will introduce a variable $\beta$ defined as $\beta=\frac{1}{1-\alpha}$, so that $\alpha=1-\frac{1}{\beta} \rightarrow 1$ as $\beta \rightarrow \infty$.
Now we rewrite equation (\ref{eq:recurRH}) as
\begin{equation}\label{eq:Deltaqi}
\Delta  \ttail{q}_i = -1 + e^{-\ttail{q}_i},
\end{equation}
with $\ttail{q}_1=\ln{\beta}$.
This equation is of the form
\begin{equation}\label{eq:Deltaqif}
\Delta  \ttail{q}_i = f(\ttail{q}_i),
\end{equation}
where $f$ is the function $f(x)=-1+e^{-x}$.
This recurrence equation seems very hard to solve exactly, but we will be able to obtain useful information about its solution by studying instead the differential equation
\begin{equation}\label{eq:diffQ}
Q'(x)=f(Q(x))
\end{equation}
with the same initial condition $Q(1)=\ln{\beta}$.
The solution to this equation is
\begin{equation}
\label{laQ}
Q(x)=\ln{(\beta-1+e^{x-1})}-x+1.
\end{equation}

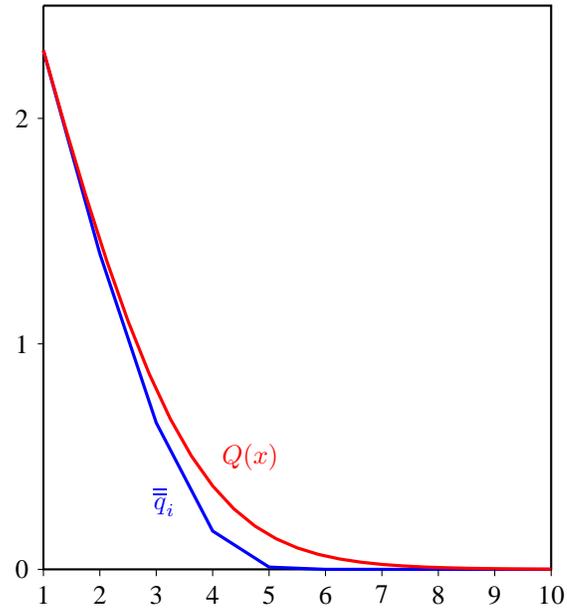
\begin{figure}[htbp]
\begin{center}
\begin{tikzpicture}[yscale=3,xscale=0.75]
\draw [thick] (1,2.5) -- (1,0) -- (10,0) -- (10,2.5) -- (1,2.5);
\draw [very thick, blue] (1,2.30) -- (2,1.40) -- (3,0.65) -- (4,0.17) -- (5,0.01) -- (6,0.00) -- (7,0.00) -- (8,0.00) -- (9,-0.00) -- (10,0.00);
\draw [very thick, red, domain=1:10] plot(\x, {ln(10-1+exp(\x-1))-\x+1}); 
\draw (0.9,0) node [left] {0} -- (1,0);
\draw (0.9,1) node [left] {1} -- (1,1);
\draw (0.9,2) node [left] {2} -- (1,2);
\draw (1,-0.03) node [below] {1} -- (1,0);
\draw (2,-0.03) node [below] {2} -- (2,0);
\draw (3,-0.03) node [below] {3} -- (3,0);
\draw (4,-0.03) node [below] {4} -- (4,0);
\draw (5,-0.03) node [below] {5} -- (5,0);
\draw (6,-0.03) node [below] {6} -- (6,0);
\draw (7,-0.03) node [below] {7} -- (7,0);
\draw (8,-0.03) node [below] {8} -- (8,0);
\draw (9,-0.03) node [below] {9} -- (9,0);
\draw (10,-0.03) node [below] {10} -- (10,0);
\node [right,red] at (4,0.5) {$Q(x)$};
\node [left,blue] at (3.5,0.3) {$\ttail{q}_i$};
\end{tikzpicture}
\end{center}
\caption{Comparison of $\ttail{q}_i$ and $Q(x)$ for $\beta=10$}
\label{plotqQ}
\end{figure}

Figure \ref{plotqQ} compares the solution $\ttail{q}_i$ (polygonal line) of recurrence equation (\ref{eq:Deltaqif}) to the solution $Q(x)$ (smooth line) of differential equation (\ref{eq:diffQ}). This plot suggests that $Q(i)$ is an upper bound for $\ttail{q}_i$. This is true, and will follow from the following lemma.

\begin{figure}[htbp]
\centering
\begin{tikzpicture}[scale=0.8]
\draw [thick] (0,8) -- (0,0) -- (9,0) -- (9,8) -- (0,8);
\draw [dotted] (0,7) -- (2,7);
\draw [dotted] (0,4) -- (2,4);
\draw [dotted] (0,3) -- (8,3);
\draw [dotted] (0,2) -- (8,2);
\draw [dotted] (2,0) -- (2,7);
\draw [dotted] (3,0) -- (3,3.833);
\draw [dotted] (8,0) -- (8,3);
\draw [very thick, blue] (2,4) -- (8,3);
\draw [very thick, red] (2,7) to [out=-80, in=120] (3,3.833) to [out=-60, in=175] (8,2);
\node [left] at (0,7) {$A(i)$};
\node [left] at (0,4) {$a_i$};
\node [left] at (0,3) {$a_{i+1}$};
\node [left] at (0,2) {$A(i+1)$};
\node [below] at (2,0) {$i$};
\node [below] at (3,-0.09) {$x$};
\node [below] at (8,0) {$i+1$};
\end{tikzpicture}
\caption{Proof of Lemma \ref{lemma:boundQ}}
\label{figlemma}
\end{figure}
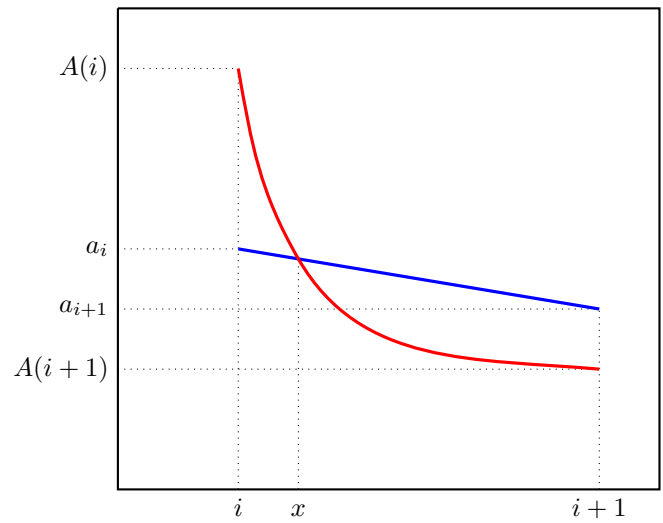

\begin{lemma}\label{lemma:boundQ}
Let $a_i$ satisfy the recurrence equation $\Delta a_{i}=f(a_i)$, and $A(x)$ satisfy the differential equation $A'(x)=f(A(x))$, where $f: [0,+\infty) \rightarrow (-\infty,0]$ is a decreasing function.
Then
\begin{equation}
A(i)\ge a_i \implies  A(i+1)\ge a_{i+1}
\end{equation}
for all $i\ge 1$.
\end{lemma}
{\em Proof\/}:
We begin by noting that both $a$ and $A$ are decreasing functions, because $f$ is negative.
Reasoning by contradiction, suppose that $A(i)\ge a_i$ but $A(i+1)<a_{i+1}$. Therefore, there exists an $x \in (i,i+1)$ such that $A(x)$ intersects the straight line joining points $(i,a_i)$ and $(i+1,a_{i+1})$, as illustrated in Figure \ref{figlemma}.
The slope of this line at $x$ is $f(a_i)$ and the slope of $A$ at point $x$ is $f(A(x))$. At the intersection we must have $f(a_i)>f(A(x))$. But $a_i>A(x)$ implies $f(a_i)<f(A(x))$, a contradiction.
\qed

\begin{corollary}
\begin{equation}
\ttail{q}_i \le Q(i) \quad \forall i \ge 1.
\end{equation}
\end{corollary}

Using this, we can rewrite equation (\ref{eq:tailV}) to obtain the following upper bound for the variance:
\begin{equation}\label{eq:boundtailV}
\sigma_{\alpha}^2 
\le \frac{2}{\alpha} \sum_{i\ge 1} Q(i) - \mu_{\alpha} - \mu_{\alpha}^2
\end{equation}

To approximate the summation, we use Euler's summation formula \cite{Knuth1},
\begin{equation}
\sum_{i\ge 1} Q(i) = \int_1^{\infty} Q(x)dx +
\sum_{k = 1}^m
\frac{B_k}{k!}(Q^{(k-1)}(\infty)-Q^{(k-1)}(1)) + R_m,
\end{equation}
where the $B_k$ are the Bernoulli numbers ($B_0=1, B_1=-\frac12, B_2=\frac16,
B_3=0, B_4=-\frac{1}{30}, \ldots$). From \cite{Knuth1} Exercise 1.2.11.2-3, we know that for even $m$, if  $Q^{(m)}(x) \geq 0$ for $x \ge 1$ then
\begin{equation}
\mid R_m\mid~\leq~\mid~\frac{B_m}{m!}(Q^{(m-1)}(\infty)-Q^{(m-1)}(1)) ~\mid.
\end{equation}
We note that, as $x \rightarrow \infty$, all derivatives of $Q(x)$ tend to zero, because they all contain the factor $f(Q(x))$, by repeated differentiation of equation (\ref{eq:diffQ}), and since $Q(\infty)=0$, we have $f(Q(\infty))=f(0)=0$.

In our case, we will apply this formula with $m=2$.
We note that
$Q(1)=\ttail{q}_1=\alpha\mu_{\alpha}$ and
$Q'(1)=f(Q(1))=f(\ttail{q}_1)=\Delta\ttail{q}_1=-\tail{q}_1=-\alpha$.
Furthermore, $Q^{(2)}(x)\ge 0$ for $x\ge 1$ because $Q'(x)=f(Q(x))$ is an increasing function. Therefore, we have
\begin{equation}
\sum_{i\ge 1} Q(i) = \int_1^{\infty} Q(x)dx +\frac12 Q(1)-\frac{1}{12} Q'(1)+R_2
\le  \int_1^{\infty} Q(x)dx + \frac12 \alpha\mu_{\alpha} + \frac16 \alpha
\end{equation}
and therefore the bound for the variance can be written as
 \begin{equation}\label{eq:boundtailV2}
\sigma_{\alpha}^2 
\le \frac{2}{\alpha}  \int_1^{\infty} Q(x)dx  + \frac{1}{3}  - \mu_{\alpha}^2
\end{equation}

Note that, until now, we have not made use of the specific form of the function $Q(x)$.
Using now formulas (\ref{laQ}) and (\ref{eq:mu}), we obtain the following upper bound for the variance:
\begin{theorem}\label{theorem:boundRH}

Under the asymptotic model for an infinite $\alpha$-full hash table with random
probing and RH collision resolution discipline, the variance of the search cost
of a random element satisfies (with $\beta = 1/(1-\alpha)$)
\begin{equation}
\sigma^2_{\alpha}\le \frac{\pi^2}{3} + \frac13 +
O\left(\frac{\ln{\beta}}{\beta}\right).
\end{equation}
\end{theorem}
\qed

This gives us an upper bound of $3.6232\ldots$ for the variance of Robin Hood Hashing.
Although a numerically computed value of approximately $1.883$ has been known for a long time, this is the first proof that this variance is bounded by a small constant as $\alpha \rightarrow 1$.
As Celis {\em et al.} observed, the fact that the variance is very small can be used to carry out a more efficient {\em mean-centered search}. If we call $X$ the random variable
``search cost of a random key''
the expected cost of this modified 
search is 
$\Theta( \mathbb{E} |X-\mu_{\alpha}|)$. But Jensen's inequality implies that

\begin{equation}
\mathbb{E} |X-\mu_{\alpha}| =
    \mathbb{E} \sqrt{(X-\mu_{\alpha})^2} \le
        \sqrt{\mathbb{E}(X-\mu_{\alpha})^2} =
             \sigma_{\alpha}
\end{equation}
so, the {\em mean value} of the search cost of a mean-centered search is proportional to the {\em standard deviation} of the cost of a standard seach.
Theorem
\ref{theorem:boundRH} then implies that this search
algorithm runs in expected constant time in a full table.

\subsection{Bounding the tail of RH}

We focus now on the tail of the distribution of the search cost, i.e. we study
\begin{equation}
\Pr\{X\ge i\} = \tail{p}_i =
\frac{1}{\alpha}\tail{q}_i =
\frac{\beta}{\beta-1}\tail{q}_i.
\end{equation}
We proved earlier that $\ttail{q}_i \le Q(i)$. By applying $f$ to both 
sides and recalling that $f$ is a decreasing function, we have
$f(\ttail{q}_i) \ge f(Q(i))$.
Using equations (\ref{eq:Deltaqif}) and (\ref{eq:diffQ}), we have
$\Delta \ttail{q}_i = -\tail{q}_i \ge Q'(i)$,
and therefore
\begin{equation}\label{eq:righttail}
\Pr\{X\ge i\} \le -\frac{\beta}{\beta-1} Q'(i) =
\frac{\beta}{\beta-1+e^{i-1}}.
\end{equation}
If we take the upper bound as the tail $\frac{\beta}{\beta-1+e^{x-1}}$ of a continuous probability function, its density function would be
\begin{equation}\label{eq:density}
p(x) = \frac{\beta e^{x-1}}{(\beta-1+e^{x-1})^2},
\end{equation}
which is symmetric around its mean (and mode) located at the point $x$ such that $e^{x-1}=\beta-1$, i.e., $x=1+\ln{(\beta-1)}$.

As a consequence, by equation (\ref{eq:righttail}), the probability that the search cost will exceed this amount by a given number of steps $k$:
\begin{equation}
\Pr\{X \ge 1+\ln{(\beta-1)+k}\} \le \frac{\beta}{\beta-1} \frac{1}{e^k+1}
                                                \rightarrow \frac{1}{e^k+1}
\end{equation}
as $\beta \rightarrow \infty$.

Therefore, as the table becomes full, the mean moves to the right without bound, but the distribution remains tightly packed to the right of the mean, and the probability that the search cost exceeds the mean by a given amount decreases exponentially with the distance.

Finally, it is interesting to note that if we shift to the left the density function (\ref{eq:density}) so it is centered around zero, we obtain
\begin{equation}
p(1+\ln{(\beta-1)}+x) = \frac{\beta}{\beta-1} \frac{e^x}{(1+e^x)^2}
\end{equation}
which, as $\beta \rightarrow \infty$, converges to $\frac{e^x}{(1+e^x)^2}$, or, equivalently, $\frac{e^{-x}}{(1+e^{-x})^2}$, the density function of a Logistic(0,1) distribution.

\section{Analysis with deletions}
\label{conborrados}

We assume a process where we first insert keys until the table reaches
load factor $\alpha$, and then we enter an infinite cycle where we alternate
one random insertion followed by one random deletion.

If the distribution of the retrieval cost is given by $p_i(\alpha)$
and a random element is inserted, the effect is described by equation
(\ref{eq:ins}).
If we then perform a random deletion, the following classical lemma\cite{feller1} shows that the distribution remains unchanged:

\begin{lemma}\label{lemma:balls}
Suppose a set contains $n$ balls of colors $1,2,\ldots,k$, such that the
probability that a ball chosen at random is of color $i$ is $p_i$.
Then, if one ball is chosen at random and discarded, the {\em a posteriori}
probability that a random ball is of color $i$ is still $p_i$.
\end{lemma}
{\em Proof\/}:
Call $p_i'$ the probability that a random ball is of color $i$
after the deletion.
The expected number of balls of color $i$ afterwards is $(n-1)p_i'$,
but that number can also be obtained as the expected number before,
$np_i$, minus the expected number of balls of color $i$ lost,
i.e.,
\begin{equation}
(n-1)p_i' = np_i - 1\cdot p_i.
\end{equation}
The result follows. \qed

Therefore, equation (\ref{eq:ins}) describes also the probability distribution
after one insert-delete step.
Now, assume the process reaches a steady state.
In that case, the distribution after the insert-delete must be equal
to the distribution before, i.e. $p_i(\alpha+\frac{1}{m}) = p_i(\alpha)$,
and replacing this in (\ref{eq:ins}) we have
\begin{equation}\label{eq:nodiffi}
p_i(\alpha) = t_i(\alpha)-t_{i+1}(\alpha).
\end{equation}
and equivalently,
\begin{equation}\label{eq:nodiffi2}
\tail{p}_i(\alpha) = t_i(\alpha).
\end{equation}
These equations play the role that equation (\ref{eq:diffi}) did for the case without deletions.
Taking tails in both sides of this equation and setting $i=1$,  we can obtain the expected search cost $\mu_{\alpha}$ as
\begin{equation}
\mu_{\alpha} = \ttail{p}_1 = \tail{t}_1 =
\frac{1}{1-\alpha},
\end{equation}
confirming the prediction that the expected successful search cost
should approach the expected {\em unsuccessful} search cost when
deletions are allowed.

For RH, from (\ref{eq:nodiffi2}) we get
$\ttail{p}_i = \tail{t}_i$,
and combining this with (\ref{eq:qiti}) we obtain
\begin{equation}
\label{eq:RHdelpi}
\ttail{p}_1 = \frac{1}{1-\alpha}, \quad
\ttail{p}_{i+1} =
	\frac{\alpha\ttail{p}_i^2}{1+\alpha\ttail{p}_i}
\end{equation}
We can use this recurrence to compute numerically the distribution for RH.

\begin{figure}[htbp]
\centering
\begin{tikzpicture}[scale=0.05]
\draw [thick] (0,100) -- (0,0) -- (100,0) -- (100,100) --(0,100);
\draw [thin] (0,0) -- (100,100);
\draw (0,0) -- (0,-3) node [below] {0};
\draw (100,0) -- (100,-3) node [below] {100};
\draw (0,0) -- (-3,0) node [left] {0};
\draw (0,100) -- (-3,100) node [left] {100};
\node [below] at (50,0) {$\beta$};
\node [red, right] at (60,50) {$\sigma^2$};
\draw [thick,red] (1, 0) -- (2, .764119604) -- (3, 1.53768652) -- (4, 2.35474566) -- (5, 3.20202070) -- (6, 4.07090868) -- (7, 4.95602766) -- (8, 5.85370062) -- (9, 6.76142610) -- (10, 7.6773737) -- (11, 8.6001731) -- (12, 9.5287720) -- (13, 10.4623453) -- (14, 11.4002320) -- (15, 12.3418980) -- (16, 13.2869258) -- (17, 14.2349116) -- (18, 15.1855618) -- (19, 16.1386132) -- (20, 17.0938388) -- (21, 18.0510458) -- (22, 19.0100642) -- (23, 19.9707422) -- (24, 20.9329522) -- (25, 21.8965762) -- (26, 22.8615134) -- (27, 23.8276698) -- (28, 24.7949602) -- (29, 25.7633122) -- (30, 26.7326598) -- (31, 27.702939) -- (32, 28.674096) -- (33, 29.646081) -- (34, 30.618843) -- (35, 31.592344) -- (36, 32.566542) -- (37, 33.541403) -- (38, 34.516888) -- (39, 35.492975) -- (40, 36.469629) -- (41, 37.446822) -- (42, 38.424559) -- (43, 39.402769) -- (44, 40.381450) -- (45, 41.360586) -- (46, 42.340152) -- (47, 43.320144) -- (48, 44.300520) -- (49, 45.281296) -- (50, 46.262428) -- (51, 47.243922) -- (52, 48.225756) -- (53, 49.207918) -- (54, 50.190394) -- (55, 51.173176) -- (56, 52.156256) -- (57, 53.139622) -- (58, 54.123258) -- (59, 55.107176) -- (60, 56.091336) -- (61, 57.075758) -- (62, 58.060410) -- (63, 59.045312) -- (64, 60.030432) -- (65, 61.015778) -- (66, 62.001322) -- (67, 62.987092) -- (68, 63.973072) -- (69, 64.959222) -- (70, 65.945570) -- (71, 66.932116) -- (72, 67.918834) -- (73, 68.905736) -- (74, 69.892806) -- (75, 70.880034) -- (76, 71.867442) -- (77, 72.854998) -- (78, 73.842702) -- (79, 74.830570) -- (80, 75.818572) -- (81, 76.806738) -- (82, 77.795030) -- (83, 78.783458) -- (84, 79.772022) -- (85, 80.760712) -- (86, 81.749536) -- (87, 82.738480) -- (88, 83.727554) -- (89, 84.716738) -- (90, 85.706040) -- (91, 86.695454) -- (92, 87.684982) -- (93, 88.674616) -- (94, 89.664370) -- (95, 90.654230) -- (96, 91.644188) -- (97, 92.634230) -- (98, 93.624384) -- (99, 94.614640) -- (100, 95.60498);
\end{tikzpicture}
\caption{The variance of RH with deletions as a function of $\beta$}
\label{plot4}
\end{figure}
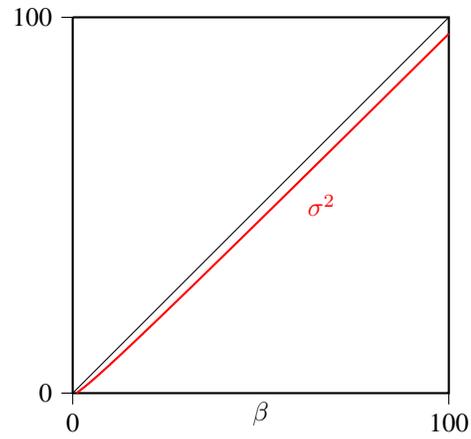

\begin{figure}[htbp]
\begin{center}
\begin{tikzpicture}[xscale=0.05,yscale=30]
\draw [thick] (1,0) -- (150,0) -- (150,0.2) -- (1,0.2) -- (1,0);
\draw [blue,thick] (1, 0.01) -- (2, 0.0099) -- (3, 0.009801) -- (4, 0.00970299) -- (5, 0.0096059601) -- (6, 0.009509900499) -- (7, 0.009414801494) -- (8, 0.009320653479) -- (9, 0.009227446944) -- (10, 0.009135172475) -- (11, 0.009043820750) -- (12, 0.008953382543) -- (13, 0.008863848717) -- (14, 0.008775210230) -- (15, 0.008687458128) -- (16, 0.008600583546) -- (17, 0.008514577711) -- (18, 0.008429431934) -- (19, 0.008345137615) -- (20, 0.008261686238) -- (21, 0.008179069376) -- (22, 0.008097278682) -- (23, 0.008016305895) -- (24, 0.007936142836) -- (25, 0.007856781408) -- (26, 0.007778213594) -- (27, 0.007700431458) -- (28, 0.007623427143) -- (29, 0.007547192872) -- (30, 0.007471720943) -- (31, 0.007397003734) -- (32, 0.007323033697) -- (33, 0.007249803360) -- (34, 0.007177305326) -- (35, 0.007105532273) -- (36, 0.007034476950) -- (37, 0.006964132181) -- (38, 0.006894490859) -- (39, 0.006825545950) -- (40, 0.006757290491) -- (41, 0.006689717586) -- (42, 0.006622820410) -- (43, 0.006556592206) -- (44, 0.006491026284) -- (45, 0.006426116021) -- (46, 0.006361854861) -- (47, 0.006298236312) -- (48, 0.006235253949) -- (49, 0.006172901409) -- (50, 0.006111172395) -- (51, 0.006050060671) -- (52, 0.005989560065) -- (53, 0.005929664464) -- (54, 0.005870367819) -- (55, 0.005811664141) -- (56, 0.005753547500) -- (57, 0.005696012025) -- (58, 0.005639051905) -- (59, 0.005582661385) -- (60, 0.005526834772) -- (61, 0.005471566424) -- (62, 0.005416850760) -- (63, 0.005362682252) -- (64, 0.005309055430) -- (65, 0.005255964875) -- (66, 0.005203405227) -- (67, 0.005151371174) -- (68, 0.005099857462) -- (69, 0.005048858888) -- (70, 0.004998370299) -- (71, 0.004948386596) -- (72, 0.004898902730) -- (73, 0.004849913703) -- (74, 0.004801414566) -- (75, 0.004753400420) -- (76, 0.004705866416) -- (77, 0.004658807752) -- (78, 0.004612219674) -- (79, 0.004566097477) -- (80, 0.004520436503) -- (81, 0.004475232138) -- (82, 0.004430479816) -- (83, 0.004386175018) -- (84, 0.004342313268) -- (85, 0.004298890135) -- (86, 0.004255901234) -- (87, 0.004213342222) -- (88, 0.004171208799) -- (89, 0.004129496711) -- (90, 0.004088201744) -- (91, 0.004047319727) -- (92, 0.004006846530) -- (93, 0.003966778064) -- (94, 0.003927110284) -- (95, 0.003887839181) -- (96, 0.003848960789) -- (97, 0.003810471181) -- (98, 0.003772366469) -- (99, 0.003734642805) -- (100, 0.003697296376) -- (101, 0.003660323413) -- (102, 0.003623720179) -- (103, 0.003587482977) -- (104, 0.003551608147) -- (105, 0.003516092066) -- (106, 0.003480931145) -- (107, 0.003446121833) -- (108, 0.003411660615) -- (109, 0.003377544009) -- (110, 0.003343768569) -- (111, 0.003310330883) -- (112, 0.003277227574) -- (113, 0.003244455299) -- (114, 0.003212010746) -- (115, 0.003179890638) -- (116, 0.003148091732) -- (117, 0.003116610814) -- (118, 0.003085444706) -- (119, 0.003054590259) -- (120, 0.003024044357) -- (121, 0.002993803913) -- (122, 0.002963865874) -- (123, 0.002934227215) -- (124, 0.002904884943) -- (125, 0.002875836094) -- (126, 0.002847077733) -- (127, 0.002818606955) -- (128, 0.002790420886) -- (129, 0.002762516677) -- (130, 0.002734891510) -- (131, 0.002707542595) -- (132, 0.002680467169) -- (133, 0.002653662497) -- (134, 0.002627125872) -- (135, 0.002600854614) -- (136, 0.002574846068) -- (137, 0.002549097607) -- (138, 0.002523606631) -- (139, 0.002498370565) -- (140, 0.002473386859) -- (141, 0.002448652990) -- (142, 0.002424166460) -- (143, 0.002399924796) -- (144, 0.002375925548) -- (145, 0.002352166292) -- (146, 0.002328644629) -- (147, 0.002305358183) -- (148, 0.002282304601) -- (149, 0.002259481555) -- (150, 0.002236886740);
\draw [thick,red] (1,0.00010) -- (2,0.00010) -- (3,0.00011) -- (4,0.00011) -- (5,0.00011) -- (6,0.00011) -- (7,0.00011) -- (8,0.00012) -- (9,0.00012) -- (10,0.00012) -- (11,0.00012) -- (12,0.00013) -- (13,0.00013) -- (14,0.00013) -- (15,0.00014) -- (16,0.00014) -- (17,0.00014) -- (18,0.00015) -- (19,0.00015) -- (20,0.00015) -- (21,0.00016) -- (22,0.00016) -- (23,0.00016) -- (24,0.00017) -- (25,0.00017) -- (26,0.00018) -- (27,0.00018) -- (28,0.00019) -- (29,0.00019) -- (30,0.00020) -- (31,0.00020) -- (32,0.00021) -- (33,0.00022) -- (34,0.00022) -- (35,0.00023) -- (36,0.00024) -- (37,0.00024) -- (38,0.00025) -- (39,0.00026) -- (40,0.00027) -- (41,0.00028) -- (42,0.00029) -- (43,0.00029) -- (44,0.00030) -- (45,0.00032) -- (46,0.00033) -- (47,0.00034) -- (48,0.00035) -- (49,0.00036) -- (50,0.00038) -- (51,0.00039) -- (52,0.00041) -- (53,0.00043) -- (54,0.00044) -- (55,0.00046) -- (56,0.00048) -- (57,0.00050) -- (58,0.00053) -- (59,0.00055) -- (60,0.00058) -- (61,0.00060) -- (62,0.00063) -- (63,0.00067) -- (64,0.00070) -- (65,0.00074) -- (66,0.00078) -- (67,0.00082) -- (68,0.00087) -- (69,0.00092) -- (70,0.00098) -- (71,0.00104) -- (72,0.00111) -- (73,0.00118) -- (74,0.00126) -- (75,0.00135) -- (76,0.00145) -- (77,0.00156) -- (78,0.00169) -- (79,0.00183) -- (80,0.00199) -- (81,0.00217) -- (82,0.00237) -- (83,0.00261) -- (84,0.00288) -- (85,0.00319) -- (86,0.00356) -- (87,0.00399) -- (88,0.00451) -- (89,0.00513) -- (90,0.00588) -- (91,0.00680) -- (92,0.00794) -- (93,0.00939) -- (94,0.01125) -- (95,0.01370) -- (96,0.01698) -- (97,0.02149) -- (98,0.02789) -- (99,0.03725) -- (100,0.05141) -- (101,0.07340) -- (102,0.10747) -- (103,0.15494) -- (104,0.19284) -- (105,0.14999) -- (106,0.04217) -- (107,0.00201) -- (108,0.00000) -- (109,0.00000) -- (110,0.00000) -- (111,0.00000) -- (112,0.00000) -- (113,0.00000) -- (114,0.00000) -- (115,0.00000) -- (116,0.00000) -- (117,0.00000) -- (118,0.00000) -- (119,0.00000) -- (120,0.00000) -- (121,0.00000) -- (122,0.00000) -- (123,0.00000) -- (124,0.00000) -- (125,0.00000) -- (126,0.00000) -- (127,0.00000) -- (128,0.00000) -- (129,0.00000) -- (130,0.00000) -- (131,0.00000) -- (132,0.00000) -- (133,0.00000) -- (134,0.00000) -- (135,0.00000) -- (136,0.00000) -- (137,0.00000) -- (138,0.00000) -- (139,0.00000) -- (140,0.00000) -- (141,0.00000) -- (142,0.00000) -- (143,0.00000) -- (144,0.00000) -- (145,0.00000) -- (146,0.00000) -- (147,0.00000) -- (148,0.00000) -- (149,0.00000) -- (150,0.00000);
\node [right,blue] at (30,0.02) {FCFS, LCFS};
\node [right,red] at (110,0.1) {RH};
\draw (-2,0) node [left] {0} -- (1,0);
\draw (-2,0.1) node [left] {0.1} -- (1,0.1);
\draw (-2,0.2) node [left] {0.2} -- (1,0.2);
\draw (1,0) -- (1,-0.005) node [below] {1};
\draw (150,0) -- (150,-0.005) node [below] {150};
\end{tikzpicture}
\end{center}
\caption{Distribution of search costs for FCFS, LCFS and RH for $\alpha=0.99$}
\label{plot2}
\end{figure}
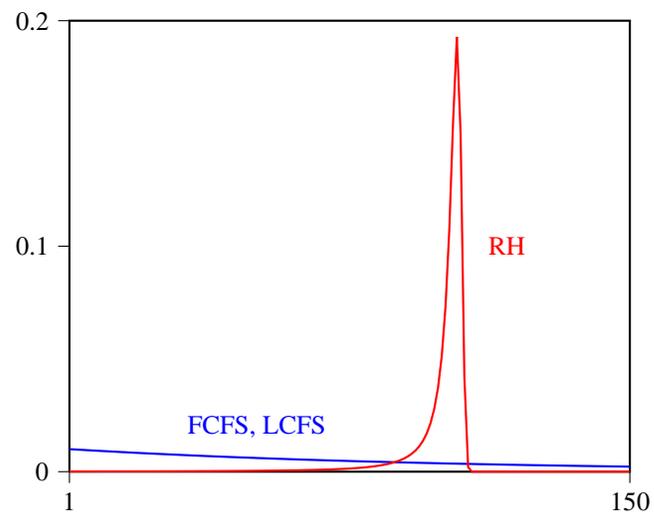

Figure \ref{plot4} shows the value of the variance of RH as a function
of $\beta=1/(1-\alpha)$, and
from the plot we may see that the variance is very close to $\beta$.
Moreover, Figure \ref{plot2} shows the distribution of the search cost for the three
methods, for $\alpha=0.99$.
As proven in \cite{GRACO} it can be seen that FCFS and LCFS are now identical
and have very large dispersion ($\sigma^2_{\alpha} =
\frac{\alpha}{(1-\alpha)^2}$), while RH retains a
much more concentrated shape. We prove that this is
indeed the case.

\subsection{Bounding the variance of RH with deletions}
We begin by rewriting the recurrence equation (\ref{eq:RHdelpi}) as
\begin{equation}\label{eq:recurqwdel}
\ttail{q}_1=\beta-1, \quad \Delta \ttail{q}_i = -\frac{\ttail{q}_i}{1+\ttail{q}_i}
\end{equation}
This equation is of the form $\Delta \ttail{q}_i = f(\ttail{q}_i)$ for $f(x)=-\frac{x}{1+x}$,
and all the conditions required in section \ref{BoundingRH} are satisfied, so we can apply the exact same technique
used there.
Solving the associated differential equation 
\begin{equation}\label{eq:diffQwdel}
Q'(x) = f(Q(x)), \quad Q(1)=\beta-1
\end{equation}
we find the solution
\begin{equation}
Q(x) = W((\beta-1)e^{\beta-x}),
\end{equation}
where $W$ is Lambert's function satisfying $x=W(x)e^{W(x)}$.
As a consequence,
proceeding as in the proof of Theorem \ref{theorem:boundRH},
we obtain the following result:
\begin{theorem}\label{theorem:boundRHwdel}
Under the asymptotic model for an infinite $\alpha$-full hash table with random
probing and RH collision resolution discipline, 
in the steady state of a sequence of insert-delete operations, the variance of the search 
cost of a random element satisfies
(with $\beta = 1/(1-\alpha)$)
\begin{equation}\label{eq:boundvar1}
\sigma^2_{\alpha} \le
\beta+\frac{1}{3}=\frac{1}{1-\alpha}+\frac{1}{3}.
\end{equation}
\end{theorem}
\qed

This proves our earlier conjecture that the variance was very close to $\frac{1}{1-\alpha}$.

\section{Acknowledgements}
We are grateful to the anonymous reviewers, for their valuable comments and suggestions, that helped us improve the paper.

\bibliography{rhvar}
\end{document}